\documentclass{appolb}
\usepackage{graphicx}
\usepackage{amssymb}
\usepackage{float}
\usepackage{hyperref}
\usepackage{url}
\usepackage{subfig}

\newcommand{\nn}{\nonumber}

\begin{document}
\title{Indication of Differential Kinetic Freeze-out at RHIC and LHC Energies%
\thanks{Presented at XI Workshop on Particle Correlations and
  Femtoscopy by R. Sahoo (Raghunath.Sahoo@cern.ch)}%
}
\author{D. Thakur, S. Tripathy, P. Garg, R. Sahoo
\address{Discipline of Physics, School of Basic Science, Indian Institute of Technology Indore, Khandwa Road, Simrol, M.P - 452020, India.}
\\
\vskip 0.8em
J. Cleymans
\address{UCT-CERN Research Centre and Department of Physics, University of Cape Town, Rondebosch 7701, South Africa}
}
\maketitle
\begin{abstract}
The transverse momentum spectra at RHIC and LHC for A+A and p+p collisions are studied with
Tsallis distributions in different approaches i.e. with and without radial flow. The information on the freeze-out
surface in terms of freeze-out volume, temperature, chemical potential and radial flow velocities
for different particle species are obtained. These parameters are found to show a systematic behavior with mass dependence. It is observed that the heavier
particles freeze-out early as compared to lighter particles and freeze-out surfaces are different for
different particles, which is a direct signature of mass dependent differential freeze-out. Further, we
observe that the radial flow velocity decreases with increasing mass. This confirms the mass ordering
behavior in collectivity observed in heavy-ion collisions. It is also observed that the systems created in peripheral heavy-ion
collisions and in proton-proton collisions are of similar thermodynamic nature.\end{abstract}
\PACS{12.40.Ee, \and 13.75.Cs, \and 13.85.-t, \and 05.70.-a}
  
\section{Introduction}

The particles produced in heavy-ion collisions carry information about the collision dynamics and evolution of the system. Therefore, the study of  particle's invariant transverse momentum ($p_T$) spectra is an important tool to understand  the system dynamics. Tsallis non-extensive statistics is commonly used to analyze the  $p_T$-spectra in hadronic and nuclear collisions at high energies. In the present work, we use different forms of Tsallis distribution \cite{Thakur} to extract the thermodynamic parameters at $\sqrt{\mathrm{s}_{NN}}$= 200 GeV and 2.76 TeV for  A+A and p+p collisions. The mass dependence of these parameters show a differential freeze-out scenario and also a similarity in p+p and A+A collisions at both energies. 


\section{Tsallis non-extensive statistics without radial flow}
The Tsallis distribution function at mid-rapidity, with finite chemical potential and without radial flow \cite{Thakur} is given by,

\begin{eqnarray}
\label{eq1}
\left.\frac{1}{p_T}\frac{d^2N}{dp_Tdy}\right|_{y=0} = \frac{gVm_T}{(2\pi)^2}
\left[1+{(q-1)}{\frac{m_T-\mu}{T}}\right]^{-\frac{q}{q-1}}
\end{eqnarray}
 where,  $~ m_T$  is the transverse mass of a particle given by $\sqrt{p_T ^2 + m^2}$, $g$ is the degeneracy and $\mu$ is the chemical potential of the system. In view of higher center of mass energies, one considers $\mu\simeq 0$, and thus the above equation modifies to~\cite{Thakur}:
\begin{eqnarray}
\label{eq2}
\left.\frac{1}{p_T}\frac{d^2N}{dp_Tdy}\right|_{y=0} = \frac{gVm_T}{(2\pi)^2}
\left[1+{(q-1)}{\frac{m_T}{T}}\right]^{-\frac{q}{q-1}}.
\end{eqnarray}

\section{Tsallis non-extensive statistics with radial flow}
In Ref.\cite{Thakur} the Tsallis distribution function has been expanded in Taylor series, after the inclusion of radial flow.
The form of the distribution up to first order in $(q-1)$ is given by
\begin{eqnarray}
\label{eq3}
\frac{1}{2\pi p_T}\frac{dN}{dp_Tdy} &&= \frac{gV}{(2\pi)^3} \nn\\ &&\biggl\{ 2 T [ r I_0(s) K_1(r) - s I_1(s) K_0(r) ]- (q-1) T r^2  I_0(s) [K_0(r)+\nn\\ &&K_2(r)]+  4(q-1)~T rs I_1(s) K_1(r)-(q-1)Ts^2 K_0(r) [I_0(s)+\nn\\ &&I_2(s)]+ \frac{(q-1)}{4}T r^3 I_0(s)  [K_3(r)+3K_1(r)]- \frac{3(q-1)}{2} T r^2 s \nn\\ &&[K_2(r)+K_0(r)] I_1(s)+ \frac{3(q-1)}{2} T s^2 r [I_0(s)+I_2(s)] K_1(r) \nn\\ &&-\left.\frac{(q-1)}{4}T s^3 [I_3(s)+3I_1(s)] K_0(r)\right\}
\end{eqnarray}

where,$~r\equiv\frac{\gamma m_T}{T} , ~s\equiv\frac{\gamma v p_T}{T}.$$~I_n(s)$ and $K_n(r)$ are  the modified Bessel functions of the first
and second kind, $V$ is the volume, $T$ 
is the Tsallis temperature, $v$ is the radial flow velocity and $q$ is the Tsallis non-extensive parameter.
For high energy collisions, the value of $q$ is $ 1 \leq q \leq 1.2$ .

\section{Results and Discussion}
 The analysis of $p_{T}$ spectra around mid rapidity is performed at $\sqrt{\mathrm{s}_{NN}}$= 200 GeV and 2.76 TeV using Eqs.\ref{eq1}, \ref{eq2} and \ref{eq3}  for  A+A and p+p collisions. It is well known that the number of binary collisions increase with number of participants. So the system created in central collisions reaches equilibrium quickly as compared to peripheral collisions. Hence, one uses non-extensive statistics while describing the thermodynamics of peripheral collisions.

\begin{figure}[ht!]
\centering
\subfloat[]{\includegraphics[scale=0.18]{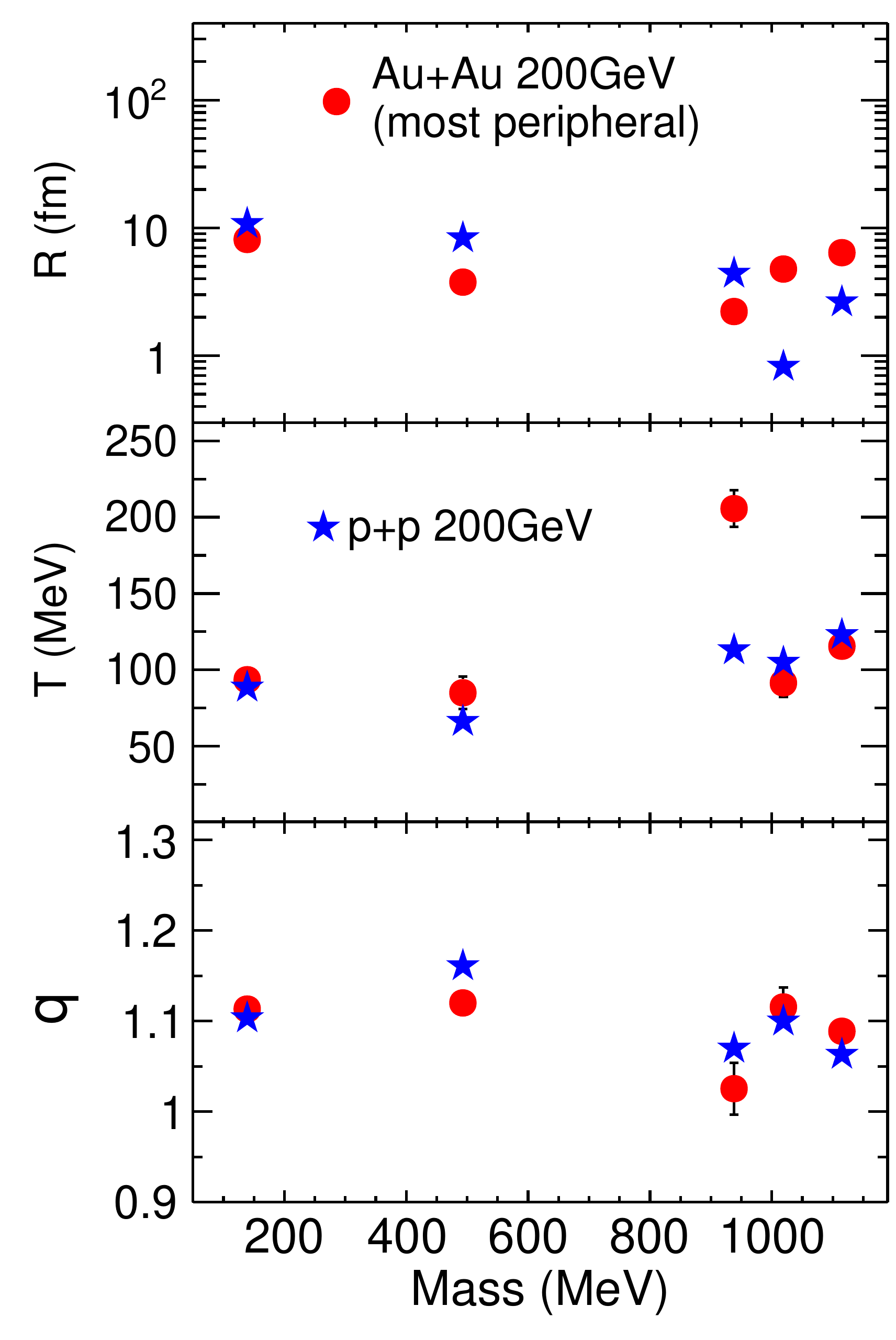}}
\subfloat[]{\includegraphics[scale=0.18]{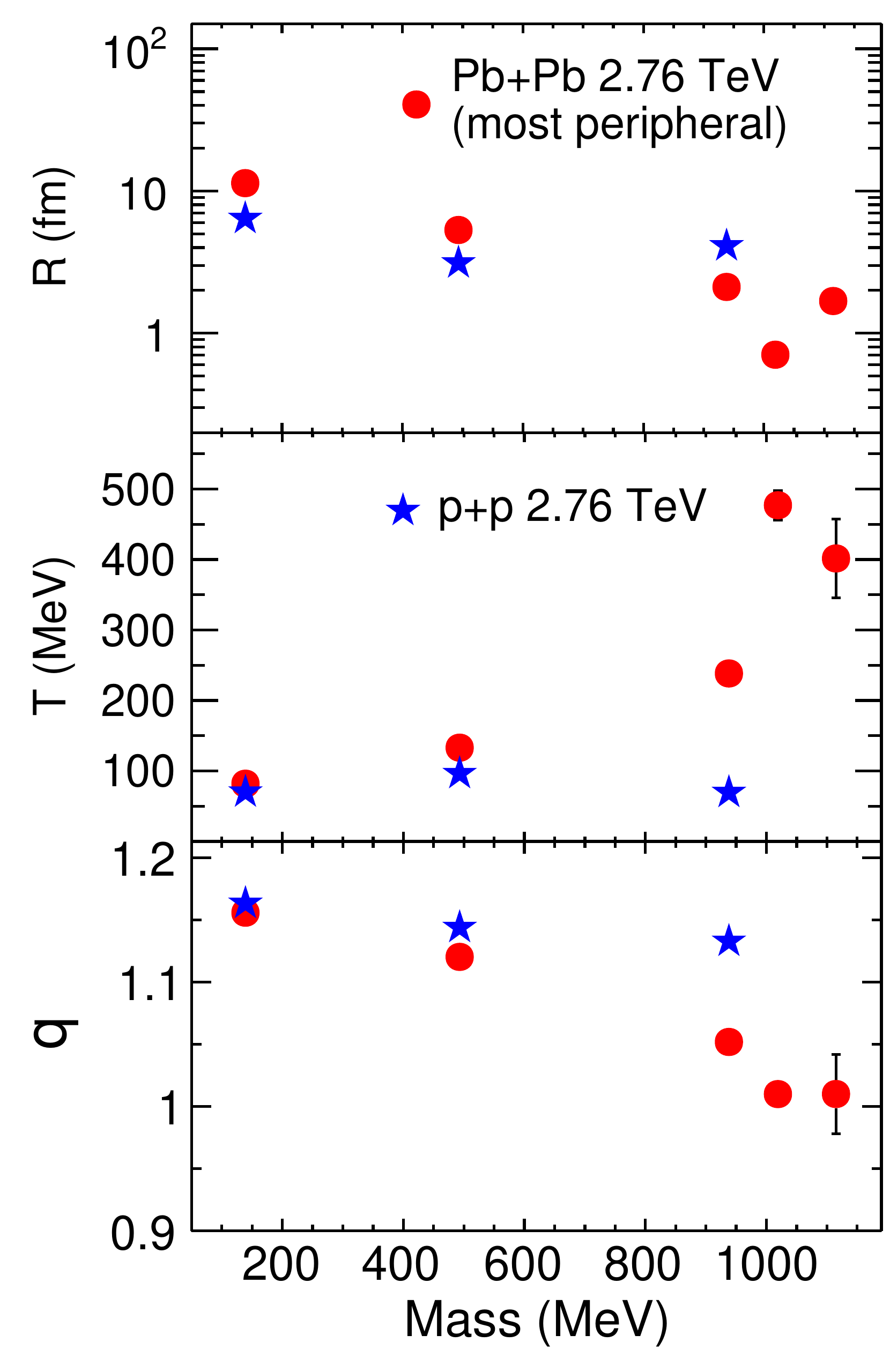}}

\caption{Mass dependence of the thermodynamic parameters using Eq.~\ref{eq2} for (a) most peripheral Au+Au
and p+p collisions at $\sqrt{s_{NN}}$ = 200 GeV (b) most peripheral Pb+Pb
and p+p collisions at $\sqrt{s_{NN}}$ = 2.76 TeV \cite{Thakur}.}
\label{Fig.1}
\end{figure}

Firstly, Eq.~\ref{eq2} is used to study the invariant $p_T$ spectra of identified particles, as shown in Fig. \ref{Fig.1}. 
It is found that the volume parameter decreases and the freeze-out temperature increases with increase in particle mass, leading to a differential freeze-out scenario. The $q$-parameter is found to be decreasing with mass for all the cases.

\begin{figure}[ht!]
\centering
\subfloat[]{\includegraphics[scale=0.25]{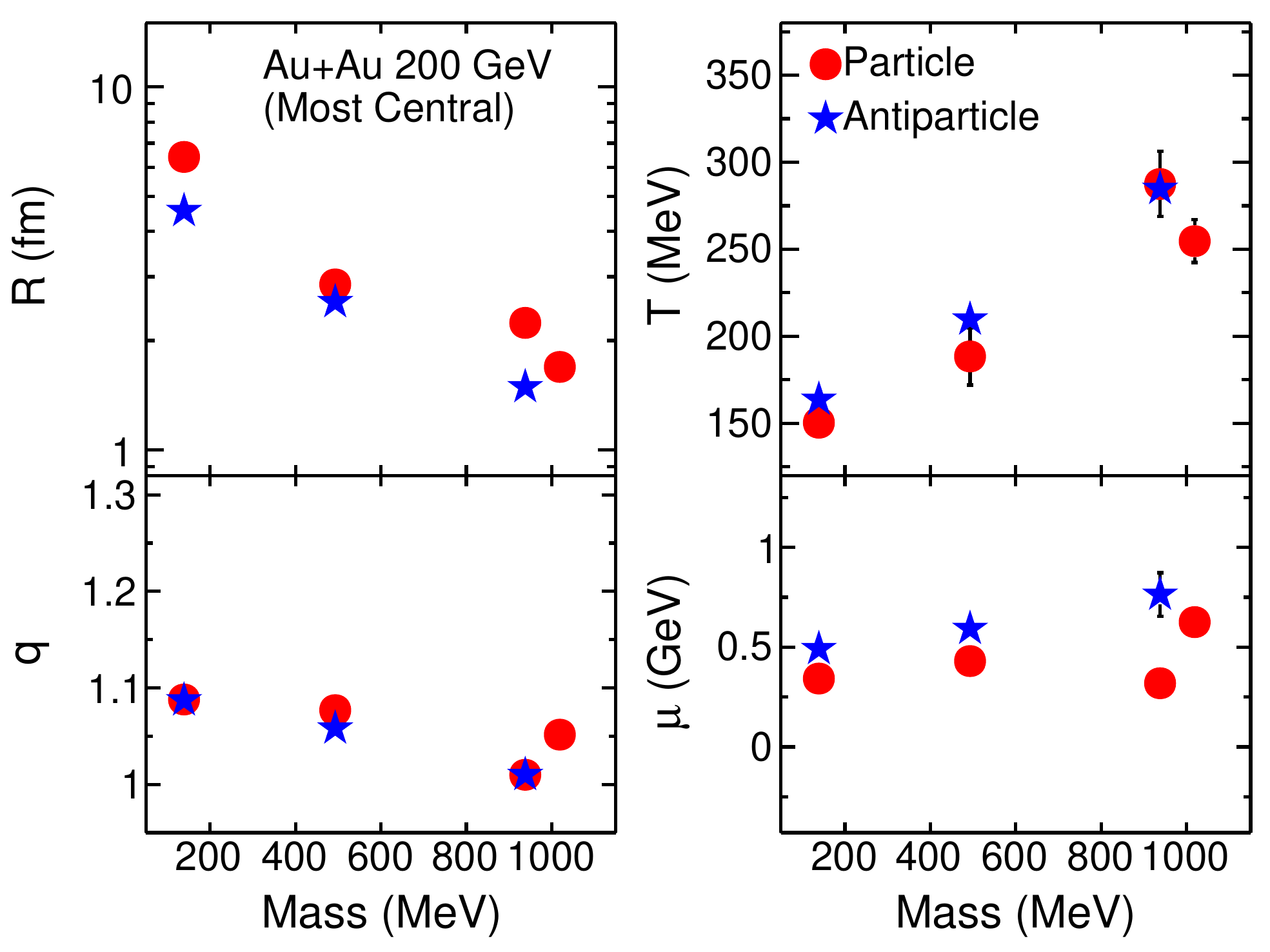}}
\subfloat[]{\includegraphics[scale=0.3]{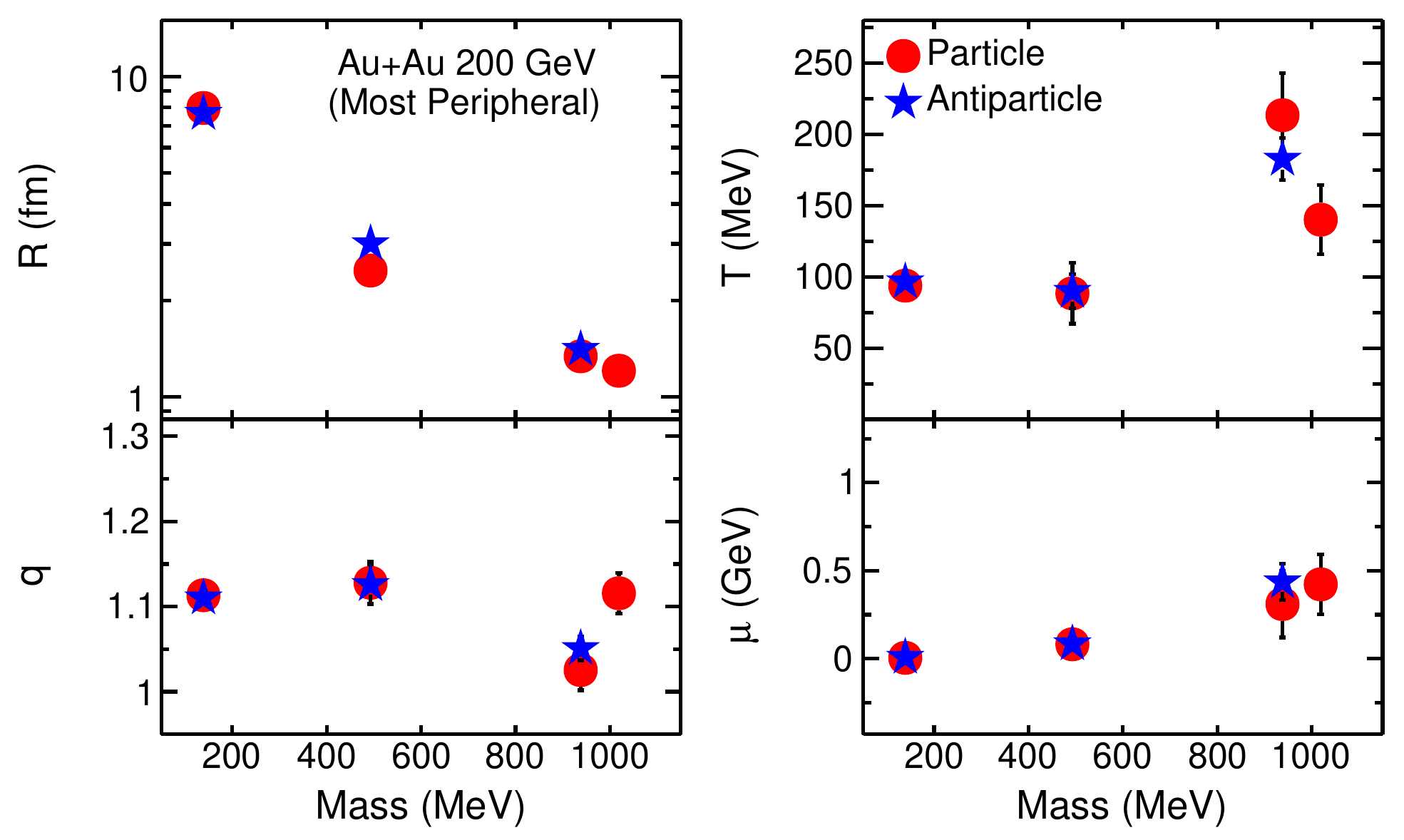}}

\subfloat[]{\includegraphics[scale=0.28]{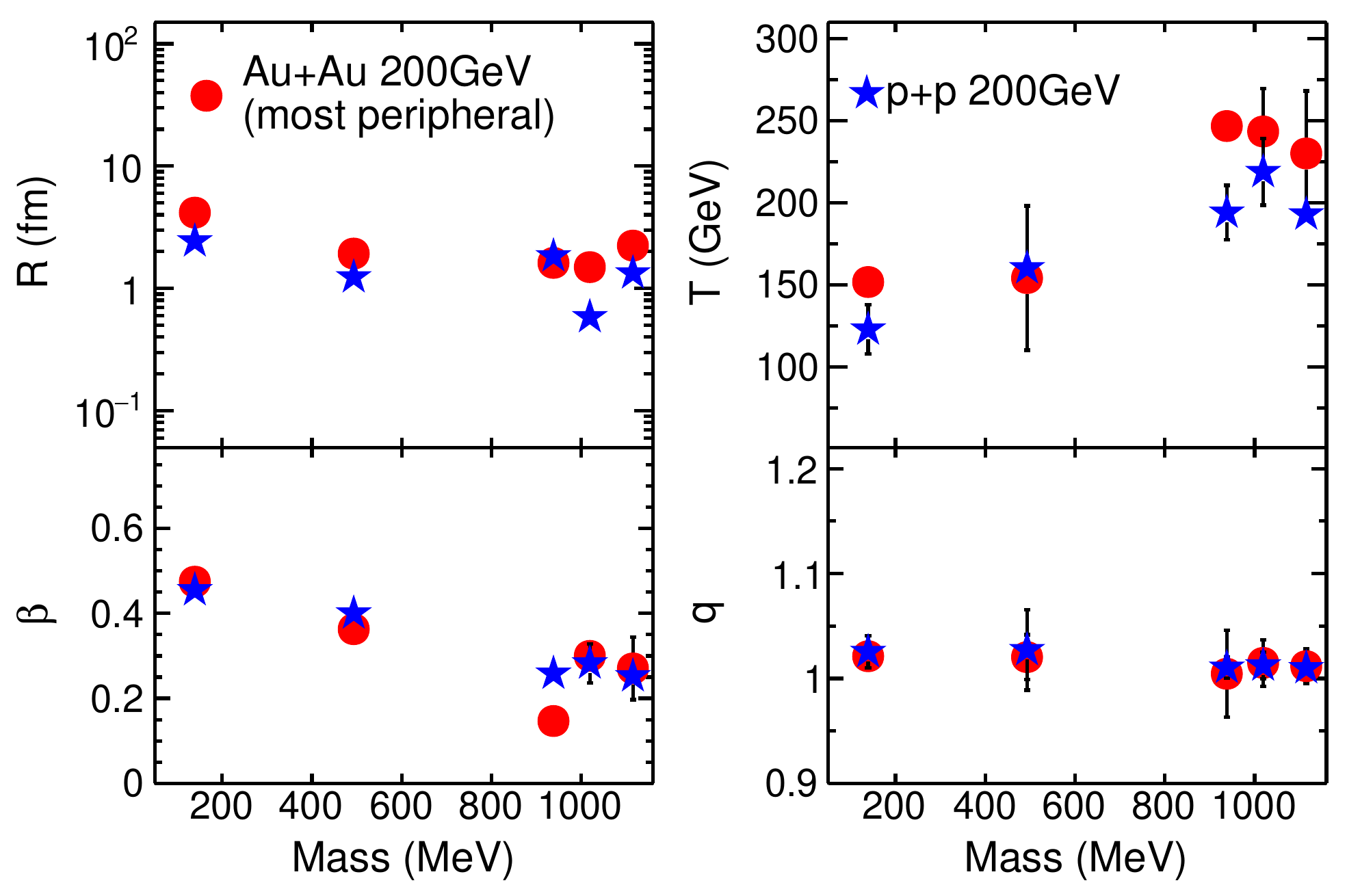}}
\subfloat[]{\includegraphics[scale=0.3]{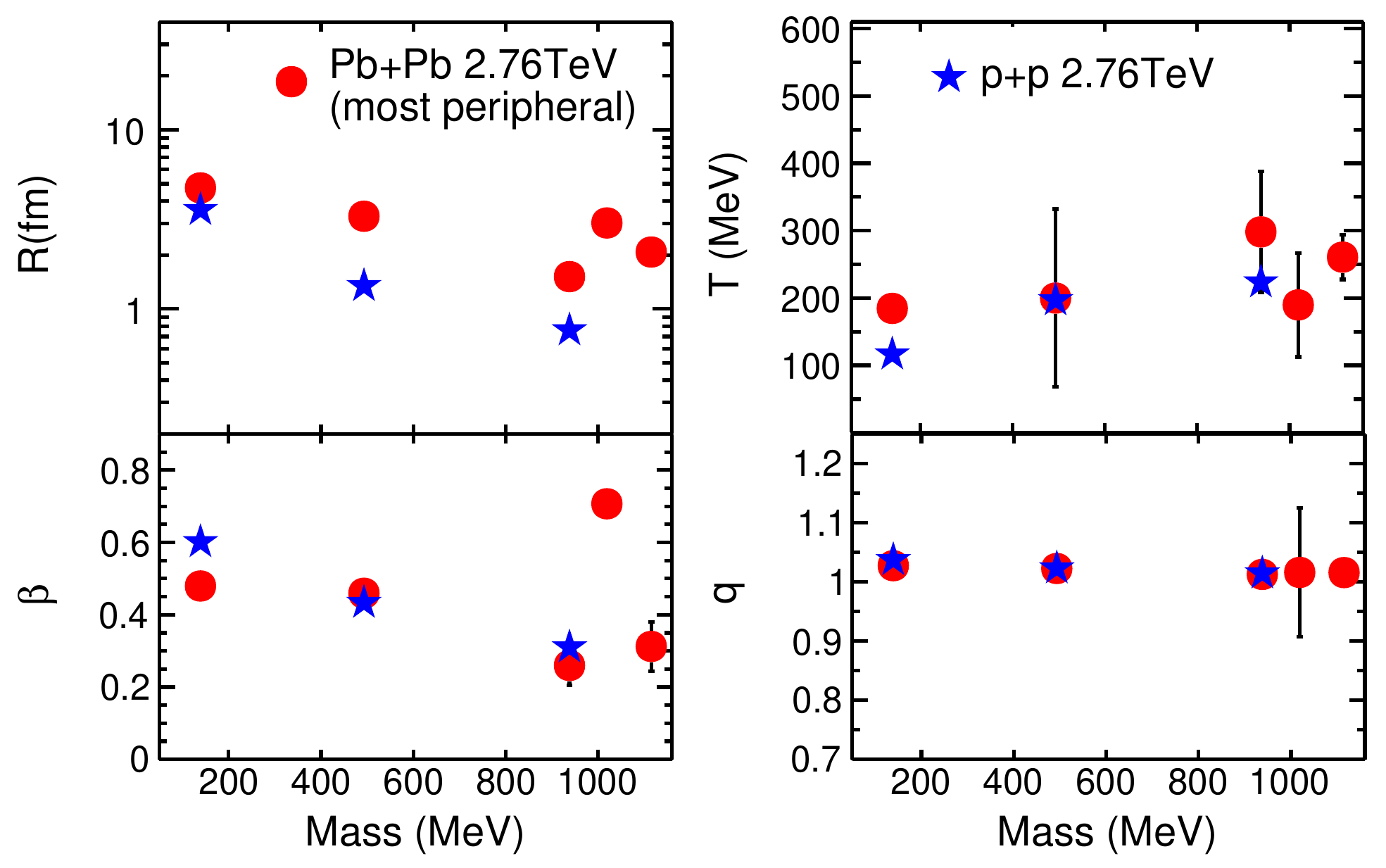}}
\caption{Mass dependent parameters using Eq.~\ref{eq1} and Eq.~\ref{eq3} for (a) most central (b) most peripheral Au+Au collisions
at $\sqrt{s_{NN}}$ = 200 GeV (c) most peripheral Au+Au
and p+p collisions at $\sqrt{s_{NN}}$ = 200 GeV (d) most peripheral Pb+Pb
and p+p collisions at $\sqrt{s_{NN}}$ = 2.76 TeV. }
\label{Fig.2}
\end{figure}

Secondly, similar procedure is followed using Eq.~\ref{eq1} for most central and peripheral Au+Au collisions at $\sqrt{\mathrm{s}_{NN}}$= 200 GeV. Similar mass dependent behavior  is depicted in Fig~\ref{Fig.2}. 
The freeze out radius parameter (R)  is not necessarily the HBT radius and also $\mu$ is not 
necessarily the chemical potential of a particle. The extracted parameters $T$ and $\mu$  (Eq.\ref{eq1}) are related by \cite{Thakur}, 
     $T = T_0  + (q-1) \mu$, 
  where $T = T_0$ for $\mu=0$. 
The parameters obtained from Eqns. \ref{eq1} (i.e. $T, \mu$ and $q$) and \ref{eq2}  (i.e. $T \equiv T_0$) satisfy this relation.

Following same procedure by using Eq.~\ref{eq3}, a similar mass dependency  is observed as shown in Fig~\ref{Fig.2}. This decrease in  $q$  value may hint that the non-extensivity is shared by  system dynamics. Here the extra parameter radial flow ($v$) is observed to decrease with particle mass. This goes inline with hydrodynamic description of the evolution of a fireball created in high-energy collisions. In summary, the above observations indicate a differential freeze-out scenario at RHIC and LHC energies. Details of the analysis could be found in Ref.\cite{Thakur}.


\begin{thebibliography}{99}

   
\bibitem{Thakur} 
  D.~Thakur, S.~Tripathy, P.~Garg, R.~Sahoo and J.~Cleymans,
  arXiv:1601.05223 [hep-ph] and references therein.
  
  
    
\end{thebibliography}
\end{document}